\documentclass[reprint,
 superscriptaddress,
 amsmath,amssymb,
 aps,
 prl,
]{revtex4-2}

	\usepackage[T1]{fontenc}
	\usepackage[utf8]{inputenc}
    \usepackage[pdftex,bookmarks=false]{hyperref}
    \usepackage{microtype}
    \usepackage{amssymb, amsmath}
    \usepackage{graphicx}
    \usepackage{dcolumn}
    \usepackage{bm}
    \usepackage{multirow}
    \usepackage{booktabs}
    \usepackage{xcolor}
    \usepackage{mathtools}
    \usepackage{mathptmx}
    \usepackage{upgreek}
    \usepackage{float}
    \usepackage{lipsum}
    \usepackage{siunitx}
    \usepackage{array}

    \hypersetup
    {
    	colorlinks = true, 
		citecolor = blue,
		bookmarksnumbered = true,
		urlcolor = green,
		pdfauthor = {},
		pdftitle = {},
		pdfsubject = {},
		pdfkeywords = {}
	}
	
    \DeclarePairedDelimiterX\braket[2]{\langle}{\rangle}{#1 \delimsize\vert #2}

    \graphicspath{Figures}

\begin{document}

\title{Neural-Network Force Field Backed Nested Sampling: Study of the Silicon $p-T$ Phase Diagram}
    \date{\today}
\author{N.~Unglert}
\affiliation{Institute of Materials Chemistry, TU Wien, 1060 Vienna, Austria} 

\author{J. Carrete}
\affiliation{Institute of Materials Chemistry, TU Wien, 1060 Vienna, Austria}

\author{L. B. Pártay}
\affiliation{Department of Chemistry, University of Warwick, Coventry CV4 7AL, UK}

\author{G. K. H. Madsen}
\email[Correspondence email address: ]{georg.madsen@tuwien.ac.at}%
\affiliation{Institute of Materials Chemistry, TU Wien, 1060 Vienna, Austria}

    
    \begin{abstract}
         Nested sampling is a promising method for calculating phase diagrams of materials, however, the computational cost limits its applicability if ab-initio accuracy is required. In the present work, we report on the efficient use of a neural-network force field in conjunction with the nested-sampling algorithm. We train our force fields on a recently reported database of silicon structures and demonstrate our approach on the low-pressure region of the silicon pressure-temperature phase diagram between 0 and \SI{16}{GPa}. 
        The simulated phase diagram shows a good agreement with experimental results, closely reproducing the melting line. Furthermore, all of the experimentally stable structures within the investigated pressure range are also observed in our simulations.
        We point out the importance of the choice of exchange-correlation functional for the training data and show how the meta-GGA r2SCAN plays a pivotal role in achieving accurate thermodynamic behaviour using nested-sampling. We furthermore perform a detailed analysis of the exploration of the potential energy surface and highlight the critical role of a diverse training data set.

    \end{abstract}

    \maketitle

\section{Introduction}

 Nested sampling (NS) is a powerful Bayesian method that can efficiently sample high-dimensional parameter spaces.\cite{skilling_nested_2004, skilling_nested_2006}
The applications of NS in materials science have progressed steadily in the past decade. While early investigations mainly focused on simple model systems such as Lennard-Jones \cite{partay_efficient_2010} or hard sphere models \cite{partay_nested_2014}, more recent work has used embedded-atom potentials to study a variety of metallic systems, including elemental metals such as Fe, Zr, and Li \cite{partay_performance_2018, marchant_nested_2022, dorrell_pressuretemperature_2020}, as well as alloys like CuAu\cite{baldock_constant-pressure_2017,CuAu_Pastewka}, AgPd \cite{partay_nested_2021} and CuPt nano-particles\cite{CuPt_ns}.

With the emergence of efficient machine-learned force fields (MLFFs), the sampling of ab initio accuracy potential energy surfaces (PES) become affordable for NS. In this context, it has been applied in conjunction with Gaussian approximation potentials (GAPs) and Moment Tensor potentials (MTPs) to predict the thermodynamic behavior of carbon \cite{marchant_carbon_2022}, platinum \cite{kloppenburg_general-purpose_2023} and AgPd \cite{rosenbrock_machine-learned_2021} alloy, respectively. 

MLFFs use statistical learning techniques to approximate the potential energy surface of a material.\cite{unke_machine_2021} 
Unlike classical interatomic potentials, MLFFs do not require extensive parametrization. Instead they provide a highly flexible functional form that has the ability to generalize across different chemical environments. Trained on datasets obtained from ab initio calculations, MLFFs can thus capture the physics of the system on par with the underlying method. However, two critical factors are paramount for their successful application. Firstly, the diversity and representativeness of the training dataset\cite{zaverkin_exploring_2022} and secondly the quality of the selected ab-initio method.

Concerning the ab initio method, Kohn-Sham density functional theory (DFT) has been the method of choice for calculating the properties of solid-state materials. 
A crucial aspect of DFT is the exchange-correlation functional, which incorporates electron-electron interactions within the system.\cite{Becke_JCP14} As an exact formulation of this functional is not available, various approximations are employed, and the choice of approximation significantly impacts the accuracy of DFT for a given problem. Traditionally, the suitability of exchange-correlation functionals has been assessed based on ground-state properties, such as lattice parameters, or cohesion energies.\cite{ tran_rungs_2016, kovacs_what_2022, borlido_exchange-correlation_2020} 
Despite their importance for practically relevant predictions, finite-temperature properties, such as the melting point, these have been less explored as target properties for evaluating the suitability of functionals, due to the computational complexity of obtaining them.\cite{Zhu_PRB17,Dorner_PRL18}
Using NS together with MLFF-based models to conduct an exhaustive exploration of the PES and give access to finite-temperature thermodynamic behaviour can bridge this gap, and thus open the door for a much more comprehensive evaluation of functional performance in a broader range of conditions.

Backed by our recently developed neural-network force field (NNFF) architecture\cite{montes-campos_differentiable_2022,carrete_deep_2023}, here we demonstrate this aspect in a NS study of the low-pressure silicon $p-T$ phase diagram. We show how the choice of a suitable exchange-correlation functional crucially influences the predicted melting temperature of Si over a large pressure range. 

In contrast to simple metallic systems, silicon stands out by its variability in chemical bonding. In its low-pressure allotrope, strong directional bonds lead to the characteristic tetrahedral coordination of the semiconducting cubic diamond phase. At higher pressures the system transitions to more closely packed structures like the well-known $\beta$-Sn phase. These circumstances complicate the use of classical interatomic force fields. Due to their rigid functional form these models are usually very poorly transferable and thus work only for the specific phases and properties they were designed for.\cite{michelin_transferability_2019}.

 This diversity of chemical bonding requires a diverse set of training data to be representative of the rich phase behaviour. To address this, we perform a detailed analysis of the configurations explored by NS, revealing a wide range of attraction basins and regions of the potential energy surface explored during the simulation. For the training data we reevaluate the database of Bartok et. al.\cite{bartok_machine_2018}, which contains around 2475 manually curated silicon structures. We show how the database, a result of the continuous efforts to create general-purpose MLFFs, possesses the diversity and representativeness necessary to deliver accurate thermodynamic predictions from a NNFF-backed NS simulation.

\section{Methodology}

\subsection{DFT}
To assess the effect of the exchange-correlation functional, we recomputed the energies and forces of the database provided by Bartók et al. \cite{bartok_machine_2018}. For the DFT calculations the PBE and r2SCAN functionals as implemented in VASP\cite{kresse_efficient_1996,kresse_ultrasoft_1999} were used.\cite{perdew_generalized_1996,r2SCAN} 
The cutoff energy for the plane wave basis was chosen as \SI{300}{eV}.
The partial occupancies for the orbitals were determined employing Fermi smearing with a smearing parameter of \SI{0.025}{eV}.
The reciprocal space sampling was performed on a Monkhorst-Pack grid with a \textit{k}-spacing of \SI{0.3}{\angstrom^{-1}} and the energy convergence criterion was set to \SI{e-5}{eV}. For the evaluation of energy-volume curves we used a denser sampling with a $k$-spacing of \SI{0.2}{\angstrom^{-1}} and a tighter convergence criterion of \SI{e-8}{eV}. We removed one configuration from the database, where a single atom is placed in a large vacuum. 

\subsection{Neural-Network Force Field}

The simulations in this work use the \textsc{NeuralIL} architecture \cite{montes-campos_differentiable_2022,carrete_deep_2023}.
Atomic coordinates are encoded into atom-centered descriptors that are invariant with respect to global rotations and translations, with relative positions of neighbors transformed into second-generation spherical Bessel descriptors \cite{kocer_continuous_2020}.
The descriptors are fed into a ResNet-inspired \cite{he_deep_2016} model, which includes a repulsive Morse contribution to avoid unphysical behavior for short interatomic distances\cite{bichelmaier_neural-network-backed_2023}. 
The implementation uses \texttt{JAX} \cite{bradbury_jax_2018} for just-in-time compilation and automatic differentiation, and \texttt{FLAX} \cite{heek_flax_2020} for simplified model construction and parameter bookkeeping. 

In order to compute second-generation spherical Bessel descriptors for our training configurations, we rely on the minimum image convention. This means that we require the training dataset to have cells large enough to fit a sphere with the corresponding cutoff radius. However, since training datasets often contain structures with a variety of different cell sizes, we need to apply a special procedure to handle this.
We use an iterative process to generate diagonal supercells of increasing size and perform a Minkowski reduction \cite{minkowski_uber_1891} to make the cell as compact as possible. This process continues until the desired cutoff fits into the cell, at which point the cycle is stopped. The result is a set of supercell structures that conform to the cutoff parameter for the descriptor generation.
However, this set may still contain configurations with significantly varying numbers of atoms. To ensure that our \texttt{JAX}-based approach is efficient, it is important to have static array sizes. Therefore, we perform a padding procedure with ghost atoms to fill up all supercells until the number of atoms is constant. This allows us to generate second-generation spherical Bessel descriptors efficiently and accurately for all configurations in the training dataset, regardless of their cell size and number of atoms.

For the descriptor generation we consider atomic environments within a cutoff of $r_{\mathrm{cut}} = \SI{4}{\angstrom}$ and choose $n_{\mathrm{max}}=6$ \cite{montes-campos_differentiable_2022}. 
For each training we randomly split the database and use $\frac{3}{4}$ of the data for training and the rest for validation.

\subsection{Nested Sampling}

NS partitions the configuration space into a nested sequence of phase space volumes confined by surfaces of iso-likelihood. Moving from the outer shells to the higher-likelihood inner shells in this nested sequence corresponds to the transition from a high-entropy fluid phase to more ordered crystalline states. Each iteration of the NS algorithm peels off a layer of the nested sequence, resulting in a corresponding sample. This approach to sampling ensures that only thermodynamically relevant structures are sampled, ultimately enabling the calculation of the thermodynamic partition function.\cite{partay_nested_2021}.

During the NS process, a group of $K$ walkers is continuously updated by replacing the  highest-energy walker at each iteration. This replacement is carried out by performing a random walk with a cloned version of one of the remaining walkers, taking a total of $L$ steps.
In the case of a constant pressure simulation, these steps involve modifying the simulation cell, through isotropic volume changes, shear transformations, or stretching operations.
Additionally, atom steps involve modifying the positions of individual atoms. This is typically accomplished through Monte Carlo steps or moves that use atomic forces, such as Galilean Monte Carlo or Hamiltonian Monte Carlo methods \cite{skilling_galilean_2019,baldock_constant-pressure_2017}.

\begin{figure}[h!]
    \centering
    \includegraphics[width=.9\columnwidth]{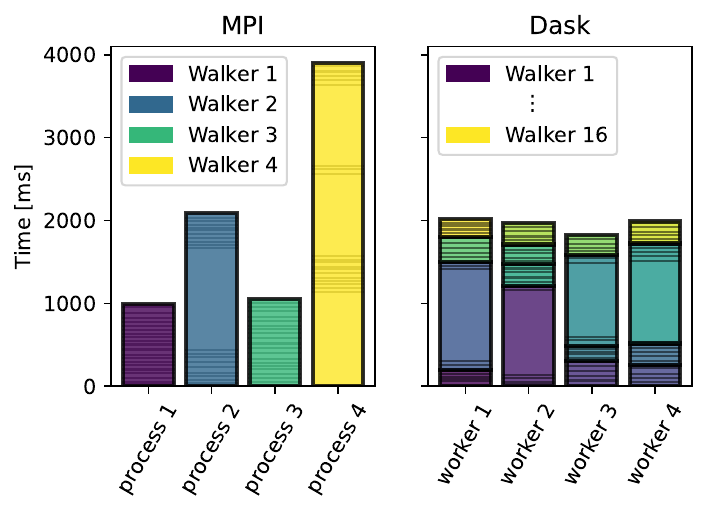}
    \caption{Synthetic scenario illustrating the advantage of the \texttt{Dask} parallelization scheme. Boxes delimited by thin lines boxes indicate individual cell or atom steps. Note, that the larger boxes correspond to the more time consuming atom move sequences. Boxes delimited by thick lines indicate whole random walks of particular walkers. For better visibility, the latter are also colored.}
    \label{fig:parallelization}
\end{figure}  
Our NS calculations were performed with a modified version of the \texttt{pymatnest} code\cite{baldock_constant-pressure_2017, partay_nested_2021}. We retained most of the logic of \texttt{pymatnest}, but adapted the parallel workflow to be managed by a scheduler provided in the Python library \texttt{Dask}.
In the original parallelization scheme\cite{baldock_constant-pressure_2017}, instead of the clone taking all the $L$ steps required to decorrelate the configuration in one iteration, each of the available $n_p$ processes is used to have $n_p$ different walkers perform $L/n_p$ steps each. 
Since we perform atom moves in sequences of several consecutive steps, and these moves involve more expensive force evaluations, they typically take significantly longer than single cell steps, which only require energy evaluations. 
As a result, depending on the random choice of step types the computational work allocated to each processor can vary substantially. In contrast, our \texttt{Dask} uses a pool of $n_w$ workers to which individual walks are dynamically assigned. The improved load balance is schematically depicted in Fig.~\ref{fig:parallelization}, showing two artificial scenarios for the original and the \texttt{Dask} parallelization. In both cases, the workload is handled by $n_{p/w}=4$ processing units and the total walk length is $L=80$ steps. In the original parallelization this corresponds to four walkers being walked for $L/n_p=20$ steps.  In this scenario, the MPI code would be required to wait for all processes to complete their random walk, resulting in significant computational overhead. Whereas, for the \texttt{Dask} example the random walk is split into $n_t=16$ slices of length $L/n_w=5$ and the workload is handled by the pool of $n_w=4$ workers. Here, $n_t$ is a parameter that should be chosen as a multiple of the number of processing units $n_w$. In the edge case of $n_t = n_w$, the \texttt{Dask} implementation becomes equivalent to the MPI scheme.

After conducting a series of convergence tests on the pristine silicon system, we chose a number of walkers of $K=600$ and a walk length of $L=1000$ steps for our simulations. With these parameters, we usually find the correct phases and the statistical spread of the transition temperatures confined to a range of \SI{200}{K}.
Samples were generated through a process of NS random walks consisting of cell and atom-movement steps. The atom movement steps were executed using the Galilean Monte Carlo algorithm in series of 8 consecutive steps. The step probability ratio for volume, stretch, shear, and atom steps was set to 2:1:1:1, respectively.
We restrict our simulation cell to a minimum aspect ratio of 0.8, to avoid pathologically thin cells forming at the early stages of the sampling.\cite{partay_nested_2021} Steps violating this constraint are discarded.
To initiate the sampling process, we first generated an initial set of 600 replicas of a cubic diamond cell with a density of \SI{2.31}{g/cm^3}. 
In a first step, we diversify the cell shapes of these structures by an initial isotropic volume scaling and a subsequent series of 1000 cell shape modifying steps. In each of these steps one shear and one stretch move is performed in random order.
In a second step, to further decorrelate the walkers, a series of 10 NS random walks each with a walk length of 100 steps was performed on the generated structures. The energy threshold for these walks was chosen to be $U_{\mathrm{initial}} = U_{\mathrm{max}} + N \cdot \SI{1}{eV}$, where $U_{\mathrm{max}}$ is the energy of the highest energy walker and $N$ is the number of atoms.

From the converged NS runs we compute the isobaric heat capacities according to
\begin{align}
    C_p = \frac{3N k_\mathrm{B}}{2} + k_\mathrm{B} 
    \beta^2 \Bigl \{ \langle Y^2 \rangle 
    - \langle Y \rangle^2 \Bigr \},
    \label{eq:heatcap}
\end{align}
where $Y$ is the microscopic enthalpy, $N$ is the number of atoms, $k_\mathrm{B}$ is the Boltzmann constant and $\beta = (k_\mathrm{B} T)^{-1}$.
The thermodynamic expectation values in equation \eqref{eq:heatcap} are evaluated using the NS partition function
\begin{align}
    \langle O \rangle = \frac{\sum_i w_i O(R_i) \exp{(-\beta Y_i)}}{\sum_i w_i \exp{(-\beta Y_i)}},
    \label{eq:expect}
\end{align}
where the sums run over all acquired samples, $w_i$ are the nested sampling weights and $O$ is an arbitrary observable depending on the configuration $R_i$. We use the isobaric heat capacity $C_p$ to locate first-order phase transitions.

\begin{figure}
    \centering
    \includegraphics[width=1.0\columnwidth]{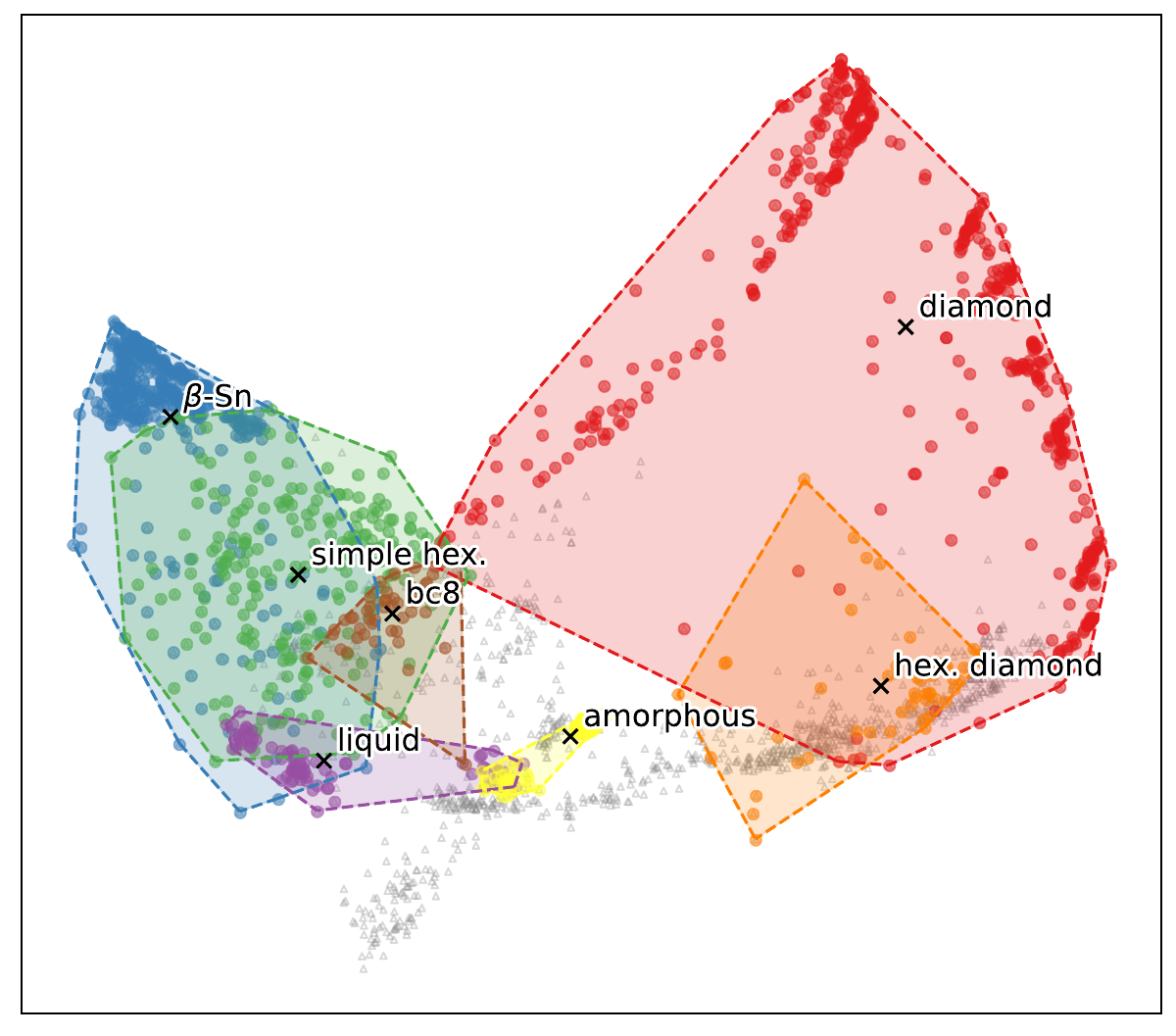}
    \caption{The first two principal components of spherical Bessel descriptors highlighting different silicon phases present in the structural database. Colored areas show the convex hull around sets of structures corresponding to a certain phase.}
    \label{fig:bartok_map}
\end{figure}


\subsection{Structure representation in 2D}

Visualizing the structural variety occurring in high-dimensional spaces, such as the 3$N_{\mathrm{atoms}}$-dimensional space of potential configurations for a system with $N_{\mathrm{atoms}}$, requires a projection into a lower dimensional space.
For that purpose, we utilize the same spherical Bessel descriptors used for encoding atomic environments for the NNFF. For a structure composed of $N_{\mathrm{atoms}}$, this results in a matrix with shape ($N_{\mathrm{atoms}}$, $n_{\mathrm{features}}$), which describes the complete structure. 
To make this representation invariant with respect to atom permutations within the structure, we compute the distributions of each of the $n_{\mathrm{features}}$ features as a histogram and divide it by the number of atoms $N_{\mathrm{atoms}}$. After flattening this yields a vector of length $n_{\mathrm{features}} n_{\mathrm{bins}}$ which is permutation invariant and independent of the system size. 
To visualize the permutation-invariant structure descriptors, we use principal component analysis (PCA) as implemented in the \texttt{scikit-learn} library \cite{pedregosa_scikit-learn_2011}. 
The histograms are calculated using $n_{\mathrm{bins}}=128$ and in a range from 0 to 4.

\subsection{Optimization and symmetry determination}

For the analysis of the walker population we perform rough relaxations of the atomic positions using our NNFF model.
For that purpose, we use the BFGS implementation contained in the \texttt{JAX} \cite{bradbury_jax_2018} library with a loose force convergence criterion of \SI{.01}{eV \angstrom^{-1}}.

For symmetry determination we employ \texttt{Spglib} \cite{togo_textttspglib_2018} as implemented in the \texttt{pymatgen} package \cite{ong_python_2013}.
Since the finite temperature structures may have slightly distorted cells, we employ a very loose symmetry precision parameter of \SI{0.3}{\angstrom} to determine the space group.

For runs that converge into strongly disordered or amorphous metastable minima, we determine the nature of the minimum by eye and by looking at the radial distribution function.

\section{Results}

\subsection{Neural-Network Force Field}

Based on the spherical Bessel descriptors the configuration space spanned by the structures in the training dataset\cite{bartok_machine_2018} is illustrated in Fig.~\ref{fig:bartok_map}.
The map is divided into several distinct regions, each representing the most prominent phases of silicon.
One of the most striking features of the map is the energetically lowest phase, cubic diamond, which occupies a significant area towards the north east. Moving towards the west, we can see the most relevant phases at intermediate pressures, such as $\beta$-Sn and simple hexagonal. 
The liquid configurations of silicon are only represented by two tiny patches located in the south west region of the map.



\begin{figure*}[t]
    \centering
    \includegraphics[width=1.\textwidth]{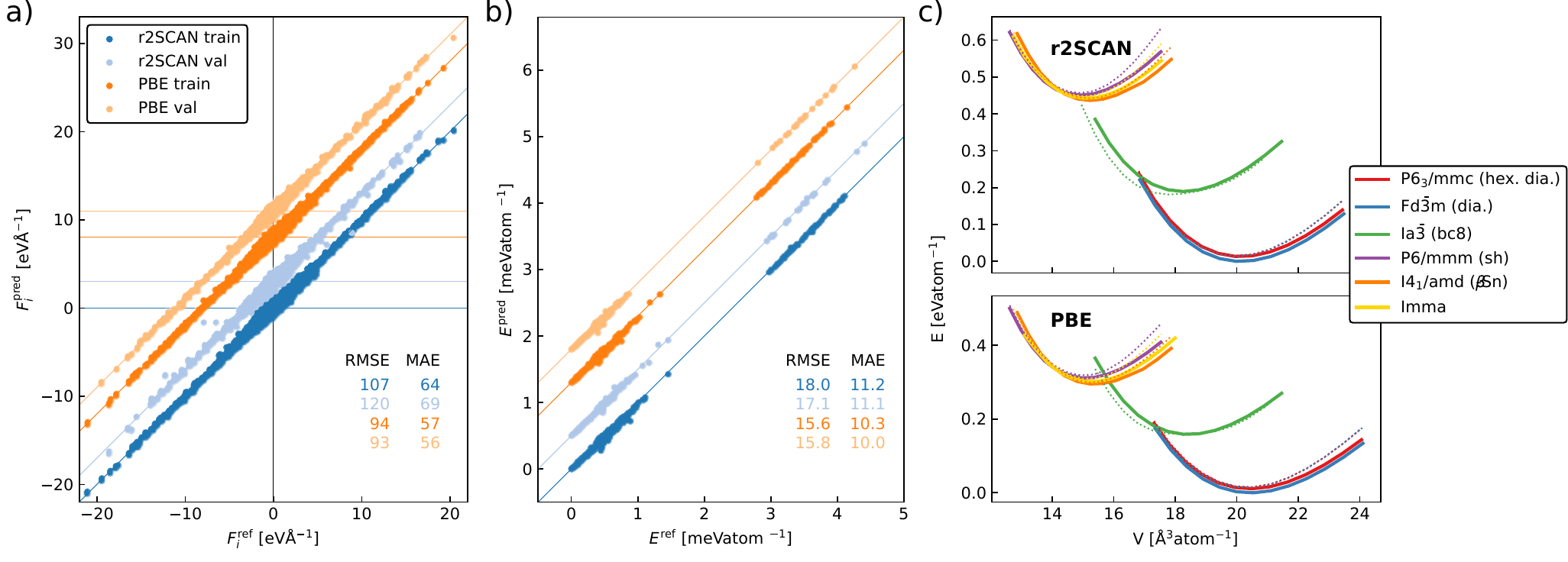}
    \caption{a) Forces and b) energy training and validation set parity plots for the NNFF.
    c) Energy-volume curves for several crystalline phases of silicon evaluated using DFT with a certain functional (solid lines) and corresponding NNFF models (dotted lines) that were trained on a database evaluated using the same functional. Top panel: r2SCAN. Bottom panel: PBE.}
    \label{fig:parity_ev}
\end{figure*}  
Figs.~\ref{fig:parity_ev}a and b show detailed parity plots as well as averaged statistics for the models trained on the r2SCAN and PBE datasets.
The energy and force errors are in the order of magnitude of \SI{10}{meV/atom} and \SI{100}{meV/\angstrom} respectively, with a slightly better result for the PBE database. Only small differences in error statistics between the training and validation sets are observed, indicating that no significant overfitting occurred in the training process.

\begin{table*}[t]
    \centering
    \setlength{\extrarowheight}{2pt} 
    \begin{tabular}{c|c|c|
    >{\centering\arraybackslash}p{0.5cm}
    >{\centering\arraybackslash}p{0.9cm}|
    >{\centering\arraybackslash}p{0.5cm}
    >{\centering\arraybackslash}p{0.9cm}|
    >{\centering\arraybackslash}p{0.5cm}
    >{\centering\arraybackslash}p{0.9cm}|
    >{\centering\arraybackslash}p{0.5cm}
    >{\centering\arraybackslash}p{0.9cm}|
    >{\centering\arraybackslash}p{0.5cm}
    >{\centering\arraybackslash}p{0.9cm}|
    >{\centering\arraybackslash}p{0.5cm}
    >{\centering\arraybackslash}p{0.9cm}|
    >{\centering\arraybackslash}p{0.5cm}
    >{\centering\arraybackslash}p{0.9cm}|
    >{\centering\arraybackslash}p{0.5cm}
    >{\centering\arraybackslash}p{0.9cm}}
        functional & size & seed & 
        \multicolumn{2}{c|}{\SI{0}{GPa}} & 
        \multicolumn{2}{c|}{\SI{4}{GPa}} & 
        \multicolumn{2}{c|}{\SI{9}{GPa}} & 
        \multicolumn{2}{c|}{\SI{10}{GPa}} & 
        \multicolumn{2}{c|}{\SI{11}{GPa}} & 
        \multicolumn{2}{c|}{\SI{12}{GPa}} & 
        \multicolumn{2}{c|}{\SI{13}{GPa}} & 
        \multicolumn{2}{c}{\SI{16}{GPa}} \\
        \hline\hline
        PBE & 16
            & 0 & \textbf{IV} & 1406 & \textbf{I} & 1146 & \textbf{II} & 906 & \textbf{II} & 941
            & \textbf{II} & 856 & \textbf{II} & 881 & \textbf{V} & 821 & \textbf{V} & 691 \\
        \hline
        \multirow{3}{*}{\parbox{1cm}{\centering r2SCAN}}
        & \multirow{3}{*}{\parbox{1cm}{\centering 16}}
            & 0 & \textbf{IV} & 1766 & \textbf{I} & 1611 & \textbf{I} & 1266 & \textbf{IV} & 1326
          & \textbf{II} & 1001 & \textbf{II} & 1006 & \textbf{V} & 1016 & \textbf{V} & 1001 \\
          & & 1 & \textbf{I} & 1801 & \textbf{I} & 1631 & \textbf{IV} & 1161 & \textbf{III} & 976
          & \textbf{II} & 976 & \textbf{II} & 1021 & \textbf{V} & 1036 & \textbf{V} & 1026 \\
          & & 2 & \textbf{I} & 1776 & \textbf{I} & 1621 & \textbf{I} & 1181 & \textbf{IV} & 1266
          & \textbf{II} & 971 & \textbf{II} & 1001 & \textbf{V} & 1046 & \textbf{V} & 1011 \\
        \hline
        \multirow{3}{*}{\parbox{1cm}{\centering r2SCAN}}
        & \multirow{3}{*}{\parbox{1cm}{\centering 32}}
            & 0 & \textbf{I} & 1681 & \textbf{I} & 1576 & \textbf{I} & 1166 & \textbf{II} & 916 
          & \textbf{II} & 976 & \textbf{II} & 991 & \textbf{V} & 1006 & \textbf{V} & 1041 \\
          & & 1 & \textbf{I} & 1551 & \textbf{I} & 1556 & \textbf{I} & 1241 & \textbf{I*} & 936 
          & \textbf{II} & 986 & \textbf{II} & 976 & \textbf{V} & 981 & \textbf{V} & 1016 \\
          & & 2 & \textbf{IV} & 1676 & \textbf{I} & 1541 & \textbf{I} & 1166 & \textbf{II} & 931 
          & \textbf{II} & 946 & \textbf{II} & 956 & \textbf{V} & 976 & \textbf{V} & 1006 \\
        \hline
        \multirow{3}{*}{\parbox{1cm}{\centering r2SCAN}}
         & \multirow{3}{*}{\parbox{1cm}{\centering 64}}
            & 0 & 
           \multicolumn{2}{c|}{\multirow{3}{*}{{\centering -}}}& 
           \textbf{I} &  1531 & 
           \multicolumn{2}{c|}{\multirow{3}{*}{{\centering -}}}& 
           \textbf{A} &  876 & 
           \multicolumn{2}{c|}{\multirow{3}{*}{{\centering -}}}& 
           \multicolumn{2}{c|}{\multirow{3}{*}{{\centering -}}}& 
           \multicolumn{2}{c|}{\multirow{3}{*}{{\centering -}}}& 
           \textbf{V} & 1011 
          \\
          & & 1 &  &  & \textbf{IV} &  1416 &  &  & \textbf{A} &  781 &  &  &  &  &  &  & \textbf{V} & 1011  \\
          & & 2 &  &  & \textbf{I*} &  1401 &  &  & \textbf{I*} & 861  &  &  &  &  &  &  & \textbf{V} & 1006 \\
    \end{tabular}
    \caption{Space groups of final structures the nested sampling converges to for different runs (\textbf{I}: cubic diamond $Fd\bar{3}m$, \textbf{II}: $\beta$-Sn $I4_1/amd$, \textbf{III}: bc8 $Ia\bar{3}$, \textbf{IV}: hexagonal diamond $P6_3/mmc$, \textbf{V}: simple hexagonal $P6/mmm$, \textbf{I*}: disordered cubic diamond $P\bar{1}$, \textbf{A}: amorphous).}
    \label{tab:converged_structures}
\end{table*}


To further test the transferabilty and accuracy of our trained models, we created a test set of structures that are not included in the training dataset. 
For this we extracted crystal structures of the most prominent silicon phases in the investigated pressure range from the materials project database. In addition, we added the body centered orthorhombic \textit{Imma} phase, first reported by McMahon et al. \cite{mcmahon_new_1993}, since it was not present in the materials project database. For each of these six phases, we created a set of isotropically scaled cells around the equilibrium volume and evaluated the energies by DFT and the corresponding NNFFs. 
The resulting energy-volume curves are shown in Fig.~\ref{fig:parity_ev} for both functionals. The NNFFs reproduced the respective curves with very similar performance. However, compared to the PBE energies, the r2SCAN energies exhibit a significant increase in energy differences. For example, the energy difference between the cubic diamond and beta-tin minimum is increased by almost 50\% in the case of r2SCAN.

\begin{figure}[h]
    \centering
    \includegraphics[width=0.85\columnwidth]{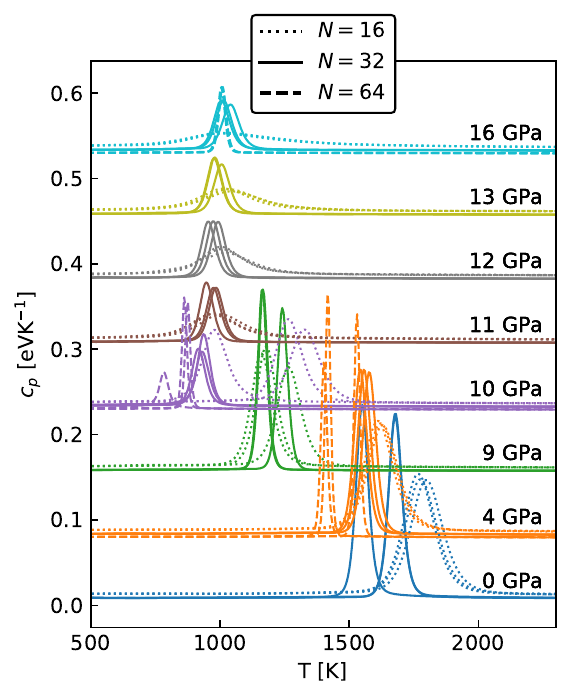}
    \caption{Heat capacities for a series of pressures in the range of 0 to \SI{16}{GPa} as calculated for system sizes of 16 (dotted lines), 32 (solid lines) and 64 atoms (dashed lines) using the r2SCAN-based model (compare Table \ref{tab:converged_structures}). For better visibility, 16- and 64-atoms heat capacities are scaled by factors of 3 and 0.3, respectively.}
    \label{fig:heat_capacities}
\end{figure}  

\subsection{Phase diagram}

We conducted NS calculations using our NNFF models across a range of eight different pressures, spanning from 0 to \SI{16}{GPa}. To account for finite size effects on the calculated quantities, we performed simulations on systems consisting of 16 and 32 silicon atoms.
Additionally, for a few specific data points, we extended the simulations to include also 64 atom systems. 
The resulting constant pressure heat capacities are shown in Fig.~\ref{fig:heat_capacities}. An overview of all calculations and the corresponding identified most stable phases is summarized in Table~\ref{tab:converged_structures}. 

Theoretically, the heat capacity diverges at first order phase transitions in the thermodynamic limit. This behavior is consistent with our findings, as the heat capacity peaks become more pronounced increasing the system size from 16 to 64 atoms. Moreover, we observe a slight shift towards higher temperatures in the smaller systems. Interestingly, this finite size effect appears to be pressure-dependent. 
Comparing the 16- and the 32-atom simulations we observe that at \SI{0}{GPa}, the deviation is more pronounced, and, as the pressure increases, the deviation gradually decreases.
For the three pressures we ran using 64 atoms, at 4 and \SI{10}{GPa} the melting temperature reduces by around \SI{100}{K}, at \SI{16}{GPa} almost no shift appears.
A similar trend was recently observed in a NS study of carbon, where the finite size effect almost diminished above a pressure of \SI{100}{GPa} \cite{marchant_carbon_2022}.

\begin{figure}[h]
    \centering
    \includegraphics[width=1.\columnwidth]{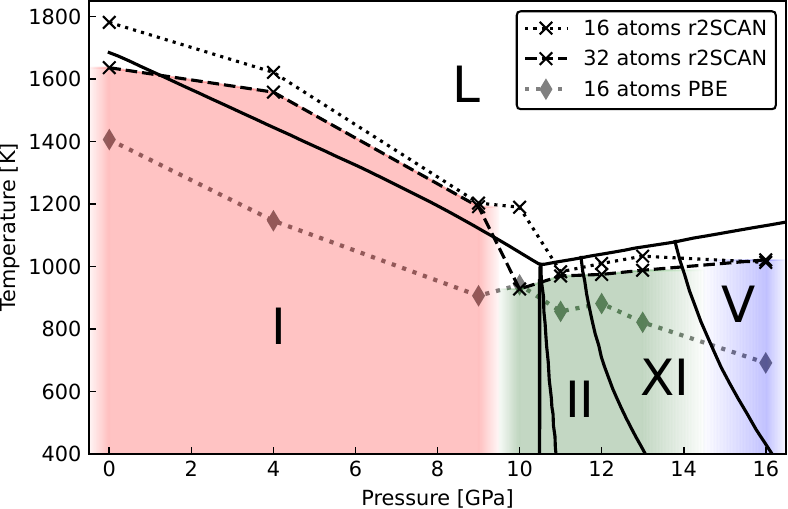}
    \caption{Pressure-temperature phase diagram of silicon. Solid black lines show the experimental phase equilibrium lines according to Ref. \cite{voronin_situ_2003} (I: cubic diamond $Fd\bar{3}m$, II: $\beta$-Sn, XI: $Imma$, V: simple hexagonal $P6/mmm$, L: liquid). Dotted and dashed black lines show 16- and 32-atom simulations using the NNFF model trained on r2SCAN data. 
    Datapoints for r2SCAN simulations represent the average of 3 independent simulations at each pressure (compare Table \ref{tab:converged_structures}).
    Grey dotted line shows the melting line of a series of 16-atom simulations using a model trained on PBE data. Colored areas show regions of stability we deduce from our runs (Red: $Fd\bar{3}m$, blue: $P6/mmm$, green: $I4_1/amd$ - $Imma$ - $P6/mmm$, see solid-solid phase transition section for detailed explanation)}
    \label{fig:phase_diagram}
\end{figure}  

Based on the calculated melting temperatures from our simulations we can construct a $p-T$ phase diagram. It is depicted in Fig.~\ref{fig:phase_diagram} together with the experimental phase diagram reported by Voronin et al. \cite{voronin_situ_2003}, which we briefly describe below.
In the low-pressure regime up to approximately \SI{10}{GPa}, the predominant phase is the cubic diamond phase and the melting line exhibits a consistent negative slope of -60~K/Pa. Increasing the pressure above 10~GPa, the $\beta$-Sn becomes stable. The cubic diamond-$\beta$-Sn-liquid triple point is found at \SI{10.5}{GPa} and \SI{1003}{K} .
Within a relatively narrow region of around \SI{2}{GPa}, the $\beta$-Sn phase occupies a distinct range and is separated from the orthorhombic \textit{Imma} phase by a phase boundary characterized by a negative slope at approximately \SI{13}{GPa}. 
As pressure increases further, the equilibrium structure transitions to a simple hexagonal phase.  
The $\beta$-Sn, \textit{Imma} and simple hexagonal phases are separated by a gentle positively sloped melting line from the the liquid phase.


The simulated r2SCAN melting temperatures (see Fig. \ref{fig:phase_diagram}, black dashed and dotted lines) reproduce the experimentally observed trends.
The 32-atom simulations show a negatively sloped melting line until \SI{10}{GPa} close to the experimental cubic diamond, $\beta$-Sn, liquid triple point.
For the higher pressures, the melting line follows the experiment with a slightly decreased slope and a small constant shift to lower temperatures.
Our simulations of different system sizes indicate that a small finite size effect even for the 64-atom runs remains for the lower pressures, while it seems to be almost absent for the higher pressure domain.
We discuss the regions of stability for the r2SCAN calculations in the following section where we perform a detailed analysis of these NS runs.

\begin{figure*}[t]
    \centering
    \includegraphics[width=1.\textwidth]{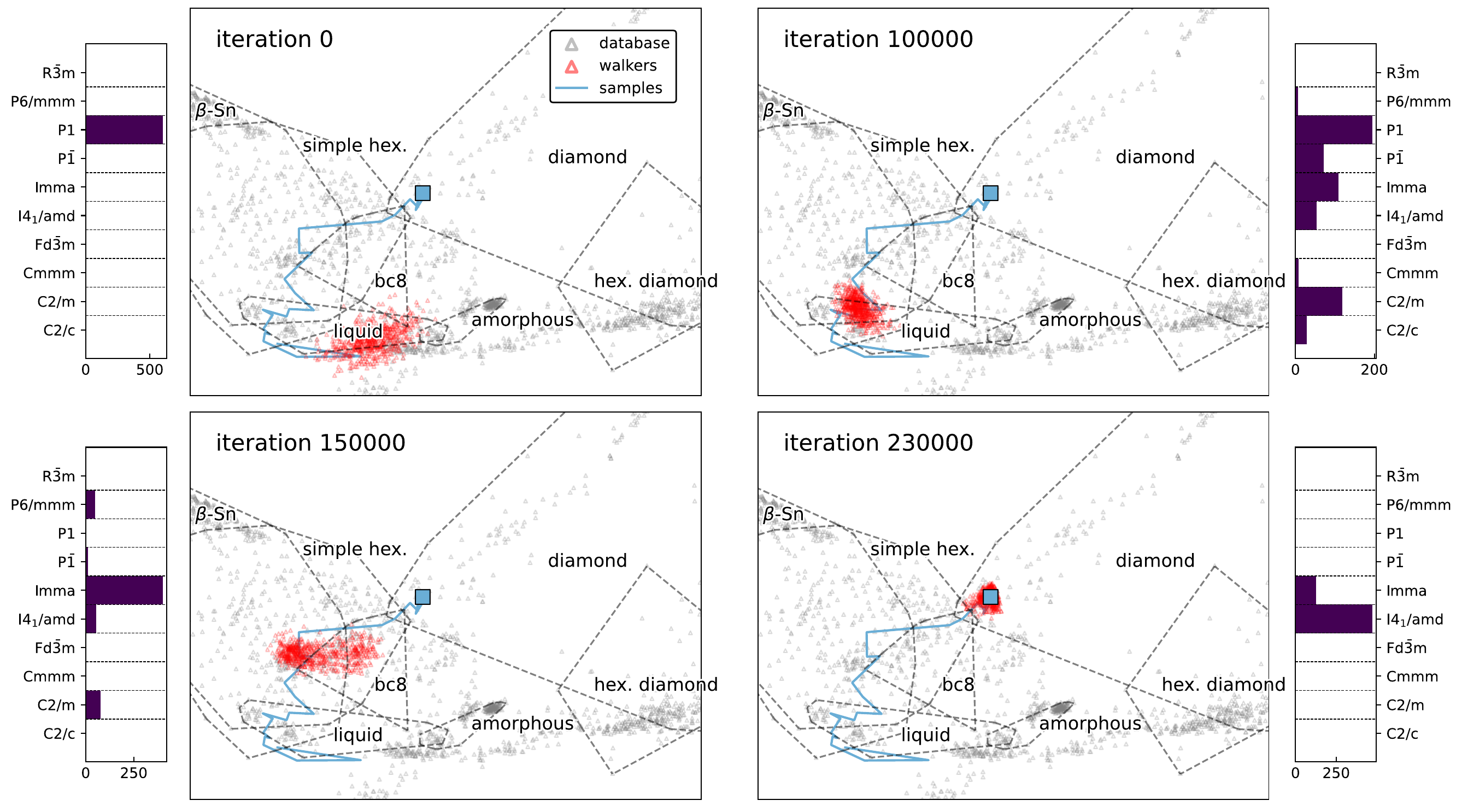}
    \caption{Evolution of the walker population of the 32-atom, seed=0, simulation at \SI{10}{GPa} in configuration space over time illustrated using the 2D configuration space map from Fig.~\ref{fig:bartok_map}. Grey points correspond to configurations in training database, dashed lines indicate convex hull of configurations belonging to a certain phase. Blue lines show the trajectory of NS samples (plotting only every 10000th sample). Red points show the walker population at the given iteration. Histograms to the side show population of space groups determined for optimized walkers.}
    \label{fig:walker_pianos}
\end{figure*}  

The 16-atom PBE simulations (see Fig. \ref{fig:phase_diagram}, grey dashed line) correctly predict the slope of the low-pressure melting line, however, they fail in the prediction of the absolute values, which are shifted to lower temperatures by approximately \SI{300}{K}. This is in line with previous ab-initio molecular dynamics simulations that predict the \SI{0}{GPa} melting temperatures to be 1687 and \SI{1450}{K} for the SCAN and the PBE functionals, respectively \cite{dorner_melting_2018}.
For higher pressures, PBE neither captures the correct trend nor the correct magnitude of the experimental melting line, with a continuing negative slope down to a transition temperature of \SI{691}{K} at \SI{16}{GPa}.
We note that these differences in melting temperatures do not arise from the algorithm finding different phases. A comparison of Table \ref{tab:converged_structures} reveals that overall similar phases are found for PBE and r2SCAN.
Instead, we relate this observation to a misprediction of the relative energies of the different silicon phases shown in Fig. \ref{fig:parity_ev}c. Due to the smaller energetic differences, the observed phase transitions can occur already at lower temperatures. In the following, we restrict our analysis to the more accurate r2SCAN results.

\subsection{Analysis of NS runs}

Fig. \ref{fig:walker_pianos} shows the evolution of the walker live set for the 32-atom NS run (seed=0) at \SI{10}{GPa} in the 2D structure representation map (compare Fig. \ref{fig:bartok_map}). 
In order to assign each of the walkers to a certain basin of the potential energy surface, we relaxed their atomic positions and determined the corresponding space group.
Initially, all walkers reside in the liquid configuration area of the training database.
At this point, the walkers are in highly disordered liquid or even gas-like states characterized by large cells and low coordination numbers.
Therefore, even after ionic relaxation, the system remains in a low-symmetry crystalline configuration, and all walkers are assigned to space group $P1$ at the start of the sampling.
As the iteration progresses, the cloud of walkers leaves the liquid area and enters the domain  spanned by the $\beta$-Sn and the simple hexagonal $P6/mmm$ phase structures of the training database.
Although the majority of walkers remain in the $P1$ space group, we observe a diverse population of other space groups, notably $Imma$, $I4_1/amd$, and $C2/m$. In subsequent snapshots, the presence of strongly disordered, liquid-like walkers diminishes. The simulation predominantly focuses on the $Imma$ phase, with the $I4_1/amd$ and $C2/m$ phases being weakly represented. Towards the end of the simulation, sampling lower enthalpy levels, a shift occurs in the population towards the $I4_1/amd$ phase, which turns out to be the most stable phase at this pressure. This contradicts the depiction in the 2D configuration space map, where the trajectory ends at the tip of the cubic diamond region, not falling into the actual $I4_1/amd$ region. We interpret this behavior as an artifact of PCA, which can not always preserve the full information from the high-dimensional space upon dimensionality reduction.
Nevertheless, the visualization in Fig. \ref{fig:walker_pianos} provides insight into how the NS algorithm explores the configuration space during the simulation. Throughout the process, the walker set encompasses a wide region in configuration space until eventually converging into the most stable basin.



\begin{figure*}[t]
    \includegraphics[width=1.\textwidth]{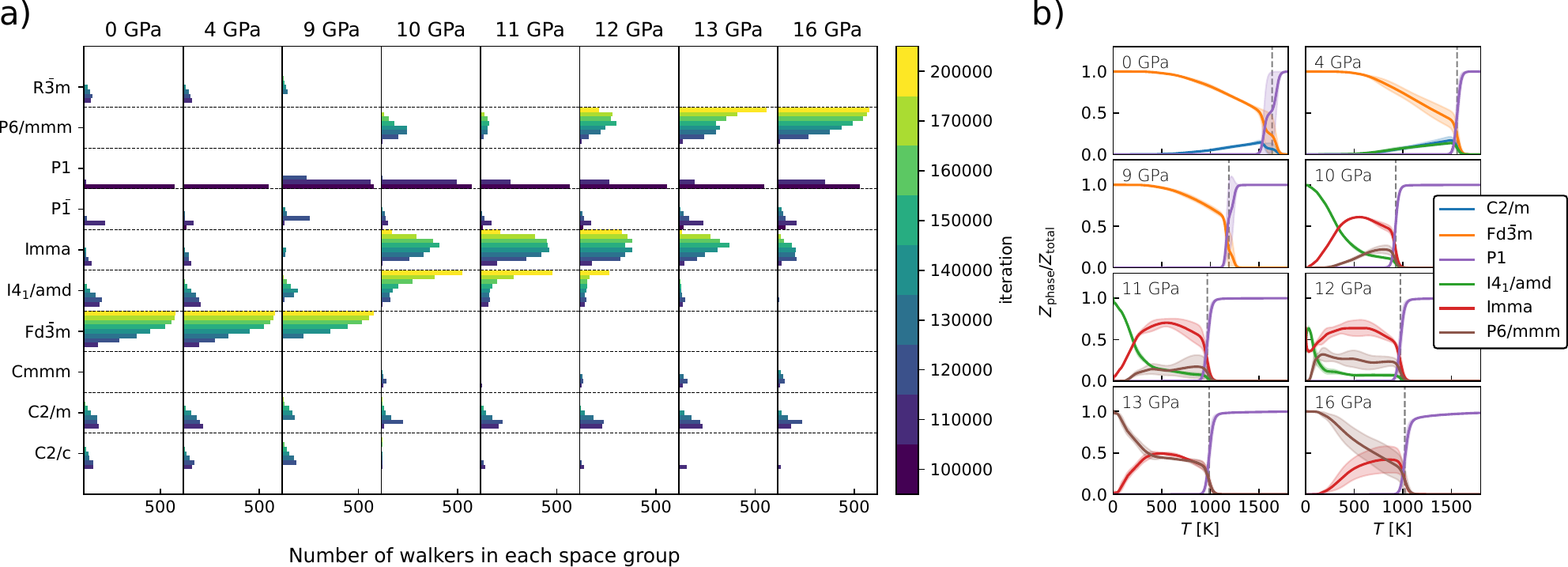}
    \caption{Analysis of the basins that are explored during the nested sampling for the 32-atom, seed=0, simulations. a) Population of the most prominent space groups over time for the walkers at each simulated pressure. Colors indicate the iteration. b) Partition function ratios of different occurring phases for all simulated pressure values, averaged over all three independent 32-atom runs, excluding two outliers discussed in the text. Colored areas show standard deviation. Dashed lines show averaged heat capacity $c_p$ peak positions (see \ref{tab:converged_structures}).}
    \label{fig:walker_optimization_partition}
\end{figure*}

A summary of the walker populations for all investigated pressures in the 32-atoms, seed=0 calculation series is presented in Figure \ref{fig:walker_optimization_partition}a. Additional analyses for the other runs can be found in the Supplementary Material.

For the three lowest pressures, a similar pattern emerges with a predominant population of the cubic diamond $Fd\bar{3}m$ phase. Although the NS algorithm visits alternative basins such as $Imma$ or $I4_1/amd$, these are quickly disregarded due to the exceptional stability of the cubic diamond phase under those conditions.
In the intermediate pressure range of 10 to \SI{12}{GPa}, two competing phases are observed. The $Imma$ phase experiences a substantial initial increase in population alongside a gradual representation of the $I4_1/amd$ phase. The $Imma$ phase later becomes depopulated towards the end of the simulation, with the $I4_1/amd$ phase emerging as the most stable. The presence of the $P6/mmm$ phase is also noted, gaining significance between 10 and \SI{12}{GPa}.
Beyond \SI{12}{GPa}, the walker population is dominated by the $P6/mmmm$ phase, which becomes the ground state. The exploration of the $I4_1/amd$ phase diminishes in importance, while the $Imma$ phase maintains a degree of population throughout the simulation.

For the pressures above \SI{10}{GPa} all r2SCAN simulations converge to the same phases consistently (see Table~\ref{tab:converged_structures}). However, discrepancies arise among the runs at lower pressures. Regardless of the system size, between 0 and \SI{9}{GPa} multiple runs converge into the hexagonal diamond $P6_3/mmc$ phase, which has been observed experimentally as a as a minor phase in indented cubic silicon.\cite{Eremenko_PSSa72} We attribute this to the small energetic difference between the actual ground state $Fd\bar{3}m$ and the $P6/mmm$ phase (see Fig.~\ref{fig:parity_ev}c). Furthermore, we observe a disordered cubic diamond phase for the \SI{4}{GPa} 64-atom simulation.
At \SI{10}{GPa}, two runs of the 16-atom simulations converge to the $P6_3/mmc$ phase, while the third run results in the $Ia\bar{3}$ space group, known as the cubic body-centered BC8 phase of silicon, which is metastable at ambient pressure \cite{voronin_situ_2003}. The BC8 phase can be obtained by slowly decompressing the metallic $\beta$-Sn phase and remains metastable unless heated above \SI{200}{\degree C} \cite{zhang_bc8_2017}. 
In the 32- and 64-atom case, we observe one simulation collapsing into a disordered cubic diamond minimum. The remaining 64-atom simulations converge to amorphous structures. We connect these discrepancies with the experimental findings of multiple metastable phases in this pressure range \cite{voronin_situ_2003}, including the BC8 phase $\mathrm{Si}_{\mathrm{III}}$ and the rhombohedral R8 phase $\mathrm{Si}_{\mathrm{III}}$. Thus, the problem becomes strongly multimodal under these conditions, hampering the sampling.

Fig.~\ref{fig:walker_pianos} furthermore gives a visual impression of
how the number of walkers determines the granularity of the potential energy surface sampling. Smaller numbers increase the likelihood of the walker cloud missing the entry point to a particular basin funnel. As a result, the set of walkers can become trapped in a metastable minimum since the Galilean Monte Carlo walk cannot traverse large energy barriers. This interpretation is supported by our comprehensive analysis of the walker populations (see Supplementary Material). In cases where a run converges to a metastable phase, the actual most stable phase is never populated, indicating that its entry point has not been found due to its small phase space volume.
The more frequent occurrence of the $P6_3/mmc$ phase in the 16-atom simulations may also be influenced by the minimum aspect ratio constraint imposed on the cell shape. We speculate that in the 16-atom case this constraint may favor the formation of the $P6_3/mmc$ phase compared to the competing $Fd\bar{3}m$ phase.

\subsection{Solid-solid phase transition}

To facilitate the identification of solid-solid phase transitions, we calculate the ratios of the contribution to the partition function by the competing phases. To achieve this, we perform optimizations on every $10^{th}$ sample obtained during the NS process and determine the corresponding space group. This enables us to assign specific samples to particular basins of the potential energy surface (PES) and separate the overall partition function into individual contributions from different phases:

\begin{align}
    Z(\beta) = \sum_i w_i e^{-\beta E_i} 
    = \sum_{i \in \mathrm{dia}} w_i e^{-\beta E_i} + \sum_{i \in \mathrm{\beta-Sn}} w_i e^{-\beta E_i} + ... 
    \label{eq:partition}
\end{align}

The result is shown as an average over the independent runs for the 32-atom simulations (excluding the two outliers at 0 and \SI{10}{GPa} discussed above) in Fig. \ref{fig:walker_optimization_partition}b. In all cases the melting transition is clearly visible in form of a sharp step in the $P1$ partition function contribution. For the pressures above \SI{9}{GPa} the competition between different phases becomes apparent. The intersections of the $I4_1/amd$ and the $Imma$ contributions at 10, 11, and \SI{12}{GPa} indicate towards a solid-solid phase transition occurring at 337, 213 and \SI{82}{K}, respectively.
Although the $Imma$ phase is significant at \SI{13}{GPa}, determining a clear phase transition point is challenging in this case.
To summarize, in the pressure range from 10 to \SI{13}{GPa}, three distinct phases, namely $I4_1/amd$, $Imma$, and $P6/mmm$, interact in a complex manner (see the green shaded area in Fig. \ref{fig:phase_diagram}).
At \SI{16}{GPa} we observe an unambiguous dominance of the $P6/mmm$ phase.

\section{Conclusion}

 In the current study, we successfully combined the nested sampling method with a fully automatically differentiable neural-network force field. By employing this powerful methodology, we can achieve ab-initio precision in our predictions, which we demonstrated by accurately simulating the pressure-temperature phase diagram of silicon. Through a comparison of the predicted melting lines from two common exchange-correlation functionals, we have demonstrated that the performance of a machine-learning model is limited by the quality of its corresponding ground truth data. Moreover, we underscore the importance of NNFF-backed nested sampling simulations, as they provide comprehensive finite temperature benchmarks for exchange correlation functionals.

Despite their great success, machine learning potentials still heavily rely on the quality, size and diversity of the training datasets to deliver accurate and reliable results. This requirement can be demanding and hinder their transferability and widespread applicability.
The inherent capability of NNFFs to handle large amounts of data facilitates the adoption of active learning methods. By developing efficient NNFF-backed nested sampling active learning approaches, we may mitigate the necessity for intricate manually curated training databases. This opens up new possibilities for purely data-driven configuration space exploration, enhancing our understanding of complex systems.

\section{Code availability}

A compatible version of \textsc{NeuralIL}, including example scripts for training and evaluation, is available on GitHub \cite{neuralil2022}. The \texttt{pymatnest} code on which our implementation is based is available on github \cite{pymatnest}.

\section{Data availability}

A dataset containing the energy and sample trajectories of all presented nested sampling calculations as well as the DFT-evaluated training databases are available on Zenodo \cite{zenodo}.

\section*{Acknowledgements}

L.B.P. acknowledges support from the EPSRC through the individual Early Career Fellowship (EP/T000163/1). J.C. and G.K.H.M. acknowledge support by the Austrian Science Fund (FWF) (SFB F81 TACO)





%

\end{document}